\documentclass[11pt]{article}
\usepackage{epstopdf}
\usepackage{color}
\usepackage{subfigure}
\usepackage{amsmath}
\usepackage{amssymb}
\usepackage{graphicx,color}
\usepackage{cite}
\usepackage{enumerate}
\usepackage{amsthm}
\usepackage{amsfonts,mathrsfs}
\usepackage{geometry}
\usepackage{ifthen}
\usepackage{graphicx}
\usepackage{float}
\usepackage{bm}
\usepackage[caption=false]{subfig}
\usepackage[utf8]{inputenc}
\usepackage{pifont}

\begin{document}

\title{\bf Tsallis relative $\alpha$ entropy of coherence dynamics in
Grover's search algorithm}

\vskip0.1in
\author{\small Linlin Ye$^1$, Zhaoqi Wu$^1$\thanks{Corresponding author. E-mail: wuzhaoqi\_conquer@163.com}, Shao-Ming
Fei$^{2}$\\
{\small\it  1. Department of Mathematics, Nanchang University,
Nanchang 330031, China}\\
{\small\it  2. School of Mathematical Sciences, Capital Normal University, Beijing 100048, China}}
\date{}
\maketitle

\noindent {\bf Abstract} {\small }\\
Quantum coherence plays a central role in Grover's search algorithm.
We study the Tsallis relative $\alpha$ entropy of coherence dynamics
of the evolved state in Grover's search algorithm. We prove that the
Tsallis relative $\alpha$ entropy of coherence decreases with the
increase of the success probability, and derive the complementarity
relations between the coherence and the success probability. We show
that the operator coherence of the first $H^{\otimes n}$ relies on
the size of the database $N$, the success probability and the target
states. Moreover, we illustrate the relationships
between coherence and entanglement of the superposition state of
targets, as well as the production and deletion of coherence in
Grover iterations.

\vskip 0.1 in


\vskip0.2 in

\noindent {\bf 1 Introduction}\\\hspace*{\fill}\\
Coherence is a fundamental property of quantum mechanics which stems
from quantum superposition principle. Quantification of coherence is
one of the most important problems in the study of quantum
coherence. Baumgratz, Gramer and Plenio \cite{TB} constituted a
rigorous framework to quantify coherence from the
viewpoint of quantum resource theories \cite{WA,CEG} which is
powerful and highly versatile. Based on this framework, some
coherence measures have been proposed \cite{SRP,MPM,YCS,WZZ}. An
alternative framework for quantifying coherence \cite{YXD} has been
formulated, and some other coherence measures \cite{ZXNZ,XCHK} have
been presented. These frameworks stimulated further researches on
relationships with other quantum resources \cite{ECM,JMB,ZXY,YYX},
coherence dynamics and related problems
\cite{QXG,DSB,SDZ,TRB,JWZ,XDY,YPYJ,AMV,CRM,WZQH,HHW,WZZL,WZZF}, and
coherence quantification in infinite-dimensional systems
\cite{JXP,YRZ}. As a significant physical resource, coherence has
diverse applications in biological systems \cite{HS,LS},
thermodynamical systems \cite{LM,MLK,VNG,MHJ,PK}, nanoscale physics
\cite{KOL}, and quantum phase transition \cite{JJ}.

Quantum algorithms may be able to solve problems that are classically
difficult. The factorization of large integers
considered to be a notoriously difficult problem on a classical
device. There is no classical factorization algorithm with
polynomial run-time. The Shor's quantum factorization algorithm
\cite{PWSP} gives a superpolynomial speedup over all known
classical factorization algorithms \cite{KEL,SDR}. Hassidim and Lloyd \cite{HAW}
proposed the Harrow-Hassidim-Lloyd (HHL) algorithm for solving linear systems of
equations, and proved that any classical algorithm generically
requires exponentially more time than the HHL algorithm.

The well-known Grover's search algorithm (GSA) has been widely used
in quantum information processing, which provides a
quadratic temporal speedup over classical search algorithms.
It has been pointed out that GSA is the repetition
of application of Grover operator $G$ \cite{GLK}, which can be
decomposed into $G=H^{\otimes n}PH^{\otimes n}O$, where $H$, $P$ and
$O$ are Hadamard operator, condition phase-shift operator and oracle
operator, respectively. The great utility of the algorithm arises
from the fact that one does not need to assume any particular
structures of the data base. As a crucial resource,
entanglement plays a significant role \cite{PR,MR,YF} in GSA. Pan,
Qiu, Mateus and Gruska \cite{MH} have shown that
the oracle operator $O$ and the reflection operator $R$ contribute
to entanglement in GSA, and demonstrated that there exists a turning
point during the Grover's iteration application.

Tsallis entropy \cite{TC} is an extension of
Shannon entropy, which plays an important role in nonextensive
statistics. Tsallis relative entropy offers an information-theoretic
basis for measuring the difference between two given distributions,
and establishes a covergence property. Its applications in the
classical system has been studied in \cite{BLP,Shiino,Tsallis}.
Quantum Tsallis relative $\alpha$ entropy \cite{AS1,AS2} is a
superior information-theoretic measure of the degree of state
purification compared with classical case, and it is monotonic under
trace preserving completely positive linear map without the
requirement that density operators are invertible \cite{FSY}.
Quantum coherence quantifier based on Tsallis relative $\alpha$
entropy has been first proposed in \cite{RAEQ}, which satisfies the
monotonicity and variational monotonicity, but not the strong
monotonicity. Zhao and Yu \cite{ZHYC} proposed a modified
well-defined quantifier for which the strong monotonicity holds. For
Tsallis relative $\alpha$ entropy, the required minimization can be
solved with an explicit answer.

Noteworthily, Tsallis relative $\alpha$ entropy of
coherence reduces to the standard relative entropy of coherence when
$\alpha\rightarrow1$, and reduces to the skew information of
coherence when $\alpha=\frac{1}{2}$, up to a constant factor. The
relative entropy of coherence can be understood as the optimal rate
for distilling a maximally coherent state from given states
\cite{WA} and has a close connection with entanglement \cite{SAS}.
The tradeoff relation of relative entropy of coherence not only
depend on the state but also accompanied by the basis-free coherence
\cite{CRM}. The skew information of coherence has an obvious
operational meaning based on the quantum metrology, and forms the
natural upper bounds for quantum correlations prepared by incoherent
operations \cite{YCS}. It characterizes the contribution of the
commutation between the density matrix of interest and the broken
observable. The skew information and $l_{1}$ norm can induce the
experimentally measurable bounds of coherence, while the relative
entropy of coherence can be exactly measured in experiment in
principle. Interestingly, Tsallis relative $\alpha$ entropy of
coherence and $l_{1}$ norm of coherence give the same ordering for
single-qubit pure states \cite{FGZ}.

The role of coherence in quantum algorithms has
attracted considerable attention in recent years
\cite{BKG,RPF,BJF,KDT,EML,MMY,CCH,BNF}. It has been found that
Deutsch-Jozsa algorithm relies on coherence during the processing,
and its precision is directly related to the recoverable coherence
\cite{HMC,JMM}. Following their footsteps, Pan and Qiu \cite{PMQ}
have explored coherence dynamics of each basic operator in GSA, and
the coherence production and depletion in terms of the $l_{1}$ norm
of coherence. Likewise, similar methods have been applied to other
algorithms, including Deutsch-Jozsa algorithm and Shor's algorithm
\cite{LYC}, and it is found that coherence depletion always exists
\cite{LYC,SHL}. Pan, Situ and Zheng \cite{MPH} have displayed the
complementarity relation between coherence and the success
probability in GSA via the $l_{1}$ norm of coherence. Coherence
determines the performance of Shor's algorithm by bounding the
success probability from below and above \cite{AFTTE}. Decoherence
in quantum algorithms has been studied \cite{BSS,HCH,SXZ,TMS,SKN},
and coherence has been explored alongside entanglement in algorithms
\cite{NMK,ANP,FCR,BSS}. The role of coherence playing in the
deterministic quantum computation with one qubit model has been
investigated explicitly \cite{FSH,MJY,WWH,GEI,KMJ}.


In this paper, we study the coherence dynamics and derive the
complementarity relations between the success probability and
coherence in GSA based on Tsallis relative $\alpha$ entropy. The
rest of the paper is organized as follows. In Section 2, we recall
GSA and Tsallis relative $\alpha$ entropy of coherence, study the
dynamics of the Tsallis relative $\alpha$ entropy of coherence in
GSA, and explore the complementarity relations between coherence and
the success probability. In Section 3, we investigate the coherence
dynamics of the state after each basic operator is applied in GSA.
In Section 4, we study the Tsallis relative $\alpha$ entropy of
coherence dynamics for different cases of the target states. By
identifying the variations before and after the basic operators are
imposed, we explore the coherence depletion and production in
Section 5. In Section 6, we compare our work with
previous works on coherence dynamics in GSA. Finally, we summarize
our results and discuss further problems in Section 7.

\vskip0.1in

\noindent {\bf 2 Tsallis relative $\alpha$ entropy of coherence in GSA and complementarity relations}\\\hspace*{\fill}\\
In Section 2.1, we first review the concepts of the Tsallis relative
$\alpha$ entropy of coherence and GSA, and then study the coherence
dynamics in GSA. We investigate the complementarity relations
between coherence and success probability in Section 2.2.
\vskip0.1in
\noindent {\bf 2.1 Tsallis relative $\alpha$ entropy of coherence in GSA}\\\hspace*{\fill}\\
The Tsallis relative $\alpha$ entropy  is defined by \cite{AS1,AS2}
\begin{equation}\label{eq1}
D_{\alpha}(\rho\|\sigma)=\frac{1}{\alpha-1}\left(f_{\alpha}(\rho,\sigma)
-1\right),~~~\alpha\in(0,1)\cup(1,\infty),
\end{equation}
where
\begin{equation*}\label{eq}
f_{\alpha}(\rho,\sigma)=\mathrm{Tr}(\rho^{\alpha}\sigma^{1-\alpha}).
\end{equation*}
It is shown that when $\alpha\rightarrow1$,
$D_{\alpha}(\rho\|\sigma)$ reduces to
$S'(\rho\parallel\sigma)=\ln2S(\rho\parallel\sigma)$, where
$S(\rho\parallel\sigma)= \mathrm{Tr}(\rho\log\rho)-
\mathrm{Tr}(\rho\log\sigma)$ is the standard relative entropy
between two quantum states $\rho$ and $\sigma$, and the logarithm
`log' is taken to be base 2. With respect to a fixed orthonormal basis
$\{|j\rangle\}_{j=1}^d$ in a $d$ dimensional Hilbert space, based on
Tsallis relative $\alpha$ entropy the coherence
$\tilde{C}_{\alpha}(\rho)$ of a density operator $\rho$ is defined
by \cite{RAEQ}
  \begin{equation}\label{eq2}
  \tilde{C}_{\alpha}(\rho)=\mathop{\mathrm{min}}\limits_{\sigma\in \mathcal{I}}
  D_{\alpha}(\rho\|\sigma)=\frac{1}{\alpha-1}\left[\left(\sum_{j=1}^{d}\langle j|\rho^{\alpha}|j\rangle^{\frac{1}{\alpha}}\right)^{\alpha}-1\right],
  \end{equation}
where $\mathcal{I}$ denotes the set of incoherent states. It is
worthwhile to note that $\tilde{C}_{\alpha}(\rho)$ does not satisfy
the strong monotonicity \cite{RAEQ}, and thus
$\tilde{C}_{\alpha}(\rho)$ is not a good coherence measure. A
well-defined coherence quantifier based on Tsallis relative $\alpha$
entropy has been presented for $\alpha\in(0,1)\cup(1,2]$
\cite{ZHYC},
  \begin{equation}\label{eq3}
  C_{\alpha}(\rho)=\mathop{\mathrm{min}}\limits_{\sigma\in \mathcal{I}}\frac{1}{\alpha-1}\left(f_{\alpha}^{\frac{1}{\alpha}}(\rho,\sigma)
  -1\right)=\frac{1}{\alpha-1}\left[\sum_{j=1}^{d}\langle j|\rho^{\alpha}|j\rangle^{\frac{1}{\alpha}}-1\right].
  \end{equation}
$C_{\alpha}(\rho)$ reduces to $\ln2 C_{r}(\rho)$ when
$\alpha\rightarrow1$, where $C_{r}(\rho)=\mathrm{Tr}(\rho\log\rho)-
\mathrm{Tr}(\rho_{\mathrm{diag}}\log\rho_{\mathrm{diag}})$ is the relative
entropy of coherence \cite{TB}. When $\alpha=\frac{1}{2}$,
$C_{\alpha}(\rho)$ reduces to $2 C_{s}(\rho)$, where
$C_{s}(\rho)=1-\sum_{j=1}^{d}\langle j|\sqrt{\rho}|j\rangle^{2}$
is the skew information of coherence \cite{YCS}.

Let $N=2^{n}$ be number of elements of the database, and $t$ the amount
of target states that meet some specific conditions in database. The
purpose of the search algorithm is to seek out the target states from database.
Grover's search algorithm makes use of a single $n$ qubit register,
starts with an $n$-qubit pure state $|0\rangle^{\otimes n}$ and applies
the Hadamard operator $H$ to get an equal superposed state,
   \begin{equation*}\label{eq}
  |\psi\rangle=\frac{1}{\sqrt{N}}\sum_{x=0}^{N-1}|x\rangle.
  \end{equation*}
It is useful to adopt the convention that $|\chi_{0}\rangle$ denotes the sum of
all not-target states $|x_{n}\rangle$, and $|\chi_{1}\rangle$ the sum of all
target states $|x_{s}\rangle$,
  \begin{equation*}\label{eq}
  |\chi_{0}\rangle=\frac{1}{\sqrt{N-t}}\sum_{x_{n}}|x_{n}\rangle,~~~
  |\chi_{1}\rangle=\frac{1}{\sqrt{t}}\sum_{x_{s}}|x_{s}\rangle.
  \end{equation*}
Simple algebra shows that $|\psi\rangle$ can be expressed as
   \begin{equation}\label{eq4}
  |\psi\rangle=\sqrt{\frac{N-t}{N}}|\chi_{0}\rangle+\sqrt{\frac{t}{N}}|\chi_{1}\rangle.
  \end{equation}
The quantum algorithm repeats the application of a quantum subroutine $G=H^{\otimes n}PH^{\otimes n}O$ named as the Grover iteration or Grover operator.
$G$ can be decomposed into four steps:

(i) Apply the oracle operator
$O=I-2\sum_{x_{s}}|x_{s}\rangle\langle x_{s}|=\sum_{x}(-1)^{f(x)}
|x\rangle\langle x|$.

(ii) Apply the Hadamard transform
$H^{\otimes n}=\frac{1}{\sqrt{N}}\sum_{x,y}(-1)^{xy}|y\rangle\langle x|$.

(iii) Perform a conditional phase shift operator
$P=2|0\rangle\langle 0|-I=\sum_{x}-(-1)^{\delta_{x0}}|x\rangle\langle x|$.

(iv) Apply the Hadamard transform $H^{\otimes n}$.

Set $\sin\theta=\sqrt{\frac{t}{N}}$ and
$\cos\theta=\sqrt{\frac{N-t}{N}}$. We have
$|\psi\rangle=\cos\theta|\chi_{0}\rangle+\sin\theta|\chi_{1}\rangle$.
After $k$ iterations of the Grover operator, the state
$|\psi\rangle$ transforms into $|\psi_{k}\rangle=\cos\theta_{k}
|\chi_{0}\rangle+\sin\theta_{k}|\chi_{1}\rangle$, where $\theta_{k}=(2k+1)\theta$. Conventionally, denote $A_{k}=\cos\theta_{k}$ and $B_{k}=\sin\theta_{k}$. One has
   \begin{equation}\label{eq5}
   |\psi_{k}\rangle=A_{k}|\chi_{0}\rangle+B_{k}|\chi_{1}\rangle.
   \end{equation}
Let $\rho_{k}=|\psi_{k}\rangle\langle\psi_{k}|$ be the density
operator of the state $|\psi_{k}\rangle$. The success probability
can be expressed as $P_{k}=\sin^{2}\theta_{k}$. Combining Eqs. (3)
with (5), we have
   \begin{equation}\label{eq6}
   C_{\alpha}(\rho_{k})=\frac{1}{\alpha-1}\left[\left(1-P_{k}\right)^{\frac{1}{\alpha}}
   \left(N-t\right)^{1-\frac{1}{\alpha}}+P_{k}^{\frac{1}{\alpha}}t^{1-\frac{1}{\alpha}}-1\right].
   \end{equation}
Under the condition $t\ll N$, we have from Eq. (6)\\
$\bullet$ when $\alpha\in(0,1)$,
   \begin{equation}\label{eq7}
   C_{\alpha}(\rho_{k})\simeq\frac{1}{\alpha-1}\left(P_{k}^{\frac{1}{\alpha}}t^{1-\frac{1}
   {\alpha}}-1\right),
   \end{equation}
$\bullet$ when $\alpha\in(1,2]$,
   \begin{equation}\label{eq8}
   C_{\alpha}(\rho_{k})\simeq\frac{N}{\alpha-1}\left(\frac{1-P_{k}}{N}\right)^{\frac{1}{\alpha}}.
   \end{equation}

{\bf Remark 2.1} Set $\alpha=\frac{1}{2}$ in Eq. (7). The skew
information of coherence of the state $|\psi_{k}\rangle$ is given by
   \begin{equation}\label{eq9}
   C_{s}(\rho_{k})\simeq1-\frac{P_{k}^{2}}{t}.
   \end{equation}
Under the limit $\alpha\rightarrow1$, from Eq. (6) the relative
entropy of coherence of the state $|\psi_{k}\rangle$ is
\begin{align}\label{eq10}
C_{r}(\rho_{k})\notag
=&\frac{1-P_{k}}{\ln2}\ln\frac{N-t}{1-P_{k}}+\frac{P_{k}}{\ln2}\ln\frac{t}{P_{k}}\\
\simeq&\frac{1-P_{k}}{\ln2}\ln\frac{N}{1-P_{k}}.
\end{align}
Here $A\simeq B$ means that $A$ asymptotically
equals to $B$ under the condition $t\ll N$. When the success
probability is increasing, according to Eqs. (7) and (8), we obtain
that the coherence is decreasing, i.e., coherence decreases with the
increase of the success probability. In particular, from Eq. (6),
when the success probability reaches 1, the coherence is
$C_{\alpha}(\rho_{k})=\frac{1}{\alpha-1}(t^{1-\frac{1}{\alpha}}-1)$.
In other words, the success probability relies on the coherence
depletion in terms of Tsallis relative $\alpha$ entropy of
coherence.

\vskip0.1in \noindent {\bf 2.2 Complementarity relations between
Tsallis relative $\alpha$ entropy of coherence and success
probability}\\\hspace*{\fill}\\
For a qubit state $|\phi\rangle=\cos\frac{\theta}{2}|0\rangle+\mathrm{e}^
{\mathrm{i}\varphi}
\sin\frac{\theta}{2}|1\rangle$, the density matrix $\rho=|\phi\rangle\langle\phi|$
has the form,
$$
\rho=\frac{1}{2}\left(\begin{matrix}
1+r_{z}\ r_{x}-\mathrm{i}r_{y}\\
r_{x}+\mathrm{i}r_{y}\ 1-r_{z}
\end{matrix}
\right),
$$
with the Bloch vector $\bm{r}=(r_{x}, r_{y}, r_{z})=
(\sin\theta\cos\varphi,\sin\theta\sin\varphi, \cos\theta)$.
The Bloch vector of the original state $|\psi\rangle=\cos\theta|\chi_{0}\rangle+\sin\theta|\chi_{1}\rangle$
in GSA is $\bm{r}(0)=(\sin2\theta, 0, \cos2\theta)$. Since $r_{y}=0$, the Grover iteration $G$ in the Bloch representation is given by
$$
G=\left[\begin{matrix}
\cos2\theta\ -\sin2\theta\\
\sin2\theta\ \cos2\theta
\end{matrix}
\right].
$$
After $k$ iterations, we have
$$
G^{k}=\left[\begin{matrix}
\cos2\theta_{k}\ -\sin2\theta_{k}\\
\sin2\theta_{k}\ \cos2\theta_{k}
\end{matrix}
\right].
$$
From
$$
\left[\begin{matrix}
r_{x}(k)\\
r_{z}(k)
\end{matrix}
\right]=G^{k}\left[\begin{matrix}
r_{x}(0)\\
r_{z}(0)
\end{matrix}
\right]=G^{k}\left[\begin{matrix}
\sin2\theta\\
\cos2\theta
\end{matrix}
\right],
$$
we obtain $r_{x}(k)\simeq-\sin2\theta_{k}$ and
$r_{z}(k) \simeq$ cos2$\theta_{k}$ when $t\ll N$.

Following the idea in \cite{MPH}, we define the
normalized $C_{\alpha}$ as
\begin{equation}\label{eq11}
N(C_{\alpha})=C_{\alpha}/C_{\alpha}^{\mathrm{max}}.
\end{equation}
Obviously, $N(C_{\alpha})\in[0, 1]$.

\noindent {\bf Theorem 1} The Tsallis relative $\alpha$ entropy of coherence and
the success probability satisfy the following complementary
relations in GSA for $t\ll N$.\\
(1) For $\alpha\in(0,1)$, it holds that
\begin{equation}\label{eq12}
N(C_{\alpha})+P_{k}^{\frac{1}{\alpha}}t^{1-\frac{1}{\alpha}}\simeq1.
\end{equation}
(2) For $\alpha\in(1,2]$, it holds that
\begin{equation}\label{eq13}
\left[N(C_{\alpha})\right]^{\alpha}+P_{k}\simeq1,
\end{equation}
where $\simeq$ means the algorithm search in large database satisfying $t\ll N$.

\noindent $\it{Proof}$. Since $r_{x}(k)\simeq-\sin2\theta_{k}$ and
$r_{z}(k)\simeq$ cos$2\theta_{k}$, simple calculation yields that
cos$^{2}\theta_{k}\simeq\frac{1}{2}(1+r_{z}(k))$ and
$\sin^{2}\theta_{k}\simeq\frac{1}{2}(1-r_{z}(k))$. According to
Eq. (6), we have
\begin{equation*}\label{eq}
C_{\alpha}(\rho_{k})=\frac{1}{\alpha-1}\left[\left(\mathrm{cos}^{2}\theta_{k}
\right)^{\frac{1}{\alpha}}
\left(N-t\right)^{1-\frac{1}{\alpha}}+\left(\sin^{2}\theta_{k}\right)^{\frac{1}
{\alpha}}t^{1-\frac{1}{\alpha}}-1\right].
\end{equation*}

When $\alpha\in(0,1)$, $t\ll N$, we obtain
\begin{equation}\label{eq14}
C_{\alpha}(\rho_{k})\simeq\frac{1}{1-\alpha}\left[1-\left(\frac{1-r_{z}
(k)}{2t}\right)^{\frac{1}{\alpha}}t\right].
\end{equation}
Since $r_{z}(k)\in [0, 1]$, the maximum of $C_{\alpha}$ is
\begin{equation}\label{eq15}
C_{\alpha}^{\mathrm{max}}\simeq\frac{1}{1-\alpha}.
\end{equation}
Then we have
\begin{equation}\label{eq16}
N(C_{\alpha})\simeq1-\left(\frac{1-r_{z}(k)}{2t}\right)^{\frac{1}{\alpha}}t.
\end{equation}
Since $P_{k}=\sin^{2}\theta_{k}\simeq\frac{1}{2}(1-r_{z}(k))$, when
$t\ll N$ and $\alpha\in(0,1)$, we get the complementarity relation (\ref{eq12}).

When $\alpha\in(1,2]$, $t\ll N$, we obtain
\begin{equation}\label{eq17}
C_{\alpha}(\rho_{k})\simeq\frac{N}{\alpha-1}\left(\frac{1+r_{z}
(k)}{2N}\right)^{\frac{1}{\alpha}}.
\end{equation}
Since $r_{z}(k)\in [0, 1]$, we have
\begin{equation}\label{eq18}
C_{\alpha}^{\mathrm{max}}\simeq\frac{N}{\alpha-1}
\left(\frac{1}{N}\right)^{\frac{1}{\alpha}},
\end{equation}
and
\begin{equation}\label{eq19}
\left[N(C_{\alpha})\right]^{\alpha}\simeq\frac{1+r_{z}(k)}{2}.
\end{equation}
As $P_{k}=\sin^{2}\theta_{k}\simeq\frac{1}{2}(1-r_{z}(k))$, when $t\ll
N$ and $\alpha\in(1,2]$, we get the complementarity relation (\ref{eq13})
between coherence and success probability.
This completes the proof. $\Box$

\noindent {\bf Remark 2.2} When $\alpha=\frac{1}{2}$ in Eq. (12), the complementarity relation between skew information of coherence and success probability is of the form,
\begin{equation*}\label{eq}
N(C_{s})+\frac{1}{t}P_{k}^{2}\simeq1.
\end{equation*}
At the limit $\alpha\rightarrow1$, Eq. (6) gives rise to
\begin{equation*}\label{eq}
C_{r}(\rho_{k})
\simeq\frac{1+r_{z}}{2\ln2}\ln\frac{2N}{1+r_{z}}.
\end{equation*}
When $r_{z}=1$, we have
\begin{equation*}\label{eq}
C_{r}^{\mathrm{max}}\simeq\frac{1}{\ln2}\ln N,~~~N(C_{r})\simeq\frac{1+r_{z}}{2}.
\end{equation*}
Since $P_{k}=\sin^{2}\theta_{k}\simeq\frac{1}{2}(1-r_{z}(k))$, we obtain the complementarity relation between relative entropy of coherence and success probability, $N(C_{r})+P_{k}\simeq1$.

According to Eqs. (12) and (13), when
$\alpha\in(0,1)$ and $P_{k}=1$, normalization of coherence satisfies
$N(C_{\alpha})+t^{1-\frac{1}{\alpha}}\simeq1$. When
$\alpha\in(1,2]$ and $P_{k}=1$, normalization of coherence satisfies
$N(C_{\alpha})\simeq0$. Noting that
$N(C_{\alpha})$ is the normalization of $C_{\alpha}$, and
$A\simeq B$ means that $A$ asymptotically equals to $B$ under the
condition $t\ll N$, this does not necessarily mean that $C_{\alpha}$
is asymptotically equals to 0, let alone it is equal to 0.

\vskip0.1in

\noindent {\bf 3 Dynamics of the Tsallis relative $\alpha$ entropy of coherence in GSA }\\\hspace*{\fill}\\
In this section, we investigate the coherence of the state after
each basic operator is applied in Grover
iteration.

Denote the first $H^{\otimes n}$ and the second $H^{\otimes
n}$ by $H_{O}$ and $H_{P}$, respectively.
In \cite{YPYJ,WA,DEJ} it has been shown that
$U=\sum_{i}\mathrm{e}^{j\alpha_{i}}|\beta(i)\rangle\langle i|$ is the general form of
any the unitary incoherent operators, where $\beta(i)$ is relabeling
of \{$i$\}. Hence, the oracle operator $O$ and the
condition phase-shift operator $P$ are incoherent operators. Moreover,
both $O$ and $P$ do not change the coherence. Denote
$|\psi_{kO}\rangle$ the state after $O$ is applied on
$|\psi_{k}\rangle$ and $\rho_{kO}=|\psi_{kO} \rangle\langle\psi_{kO}|$. We
have
   \begin{equation}\label{eq20}
   |\psi_{kO}\rangle\equiv O|\psi_{k}\rangle=A_{k}|\chi_{0}\rangle-B_{k}|\chi_{1}\rangle.
   \end{equation}
Combining Eqs. (3) with (5), we have $C_{\alpha}(\rho_{kO})=C_{\alpha}(\rho_{k})$,
where $\alpha\in(0,1)\cup(1,2]$.

In the computational basis, the $N$-dimensional Hadamard matrix has
the following three properties: (i) The common
coefficient is $\frac{1}{\sqrt{N}}$. (ii)
The elements of the first row and the first column
are 1. (iii) In any other rows or
columns, a half of the elements are 1, and the others are $-1$. Let
$H_{y,x}$ denote the element of the $y$th row and the $x$th column
in a Hadamard matrix and $t_{y}$ denote the number of $H_{y,x}=1$,
$t_{y}=|\{H_{y,x}|H_{y,x}=1,x\in{\{x_{s}\}}\}|$. Denote by
$|\psi_{kH_{O}}\rangle$ the state after $H_{O}$ is applied on
$|\psi_{kO}\rangle$ and $|\psi_{kH_{P}}\rangle$ the state after
$H_{P}$ is applied on $PH^{\otimes n}|\psi_{kO}\rangle$. We have
\begin{align}\label{eq21}
|\psi_{kH_{O}}\rangle\notag
\equiv& H^{\otimes n}|\psi_{kO}\rangle\\
=&\left[\frac{1}{\sqrt{N}}\sum_{y=0}^{N-1}\frac{A_{k}}{\sqrt{N-t}}\sum_{x\in \{x_{n}\}}
(-1)^{xy}-\frac{1}{\sqrt{N}}\sum_{y=0}^{N-1}\frac{B_{k}}{\sqrt{t}}\sum_{x\in \{x_{s}\}}
(-1)^{xy}\right]|y\rangle,
\end{align}
and
\begin{equation}\label{eq22}
|\psi_{kH_{P}}\rangle
\equiv H^{\otimes n}PH^{\otimes n} |\psi_{kO}\rangle
=|\psi_{k+1}\rangle.
\end{equation}

\noindent {\bf Theorem 2} The Tsallis relative $\alpha$ entropy of
coherence of the states $|\psi_{kH_{O}}\rangle$ and $|\psi_{kH_{P}}\rangle$ are given by
\begin{equation}\label{eq23}
C_{\alpha}(\rho_{kH_{O}})\simeq\frac{1}{\alpha-1}\left[\left(P_{k}\gamma(t,t_{y})
   \right)^{\frac{1}{\alpha}}N^{1-\frac{1}{\alpha}}-1\right],
\end{equation}
and
\begin{equation}\label{eq24}
C_{\alpha}(\rho_{kH_{P}})\simeq\frac{1}{\alpha-1}\left(P_{k+1}^{\frac{1}{\alpha}}t^{1-\frac{1}
{\alpha}}-1\right),
\end{equation}
respectively for $t\ll N$ and $\alpha\in(0,1)$, and
\begin{equation}\label{eq25}
C_{\alpha}(\rho_{kH_{O}})\simeq\frac{N}{\alpha-1}\left(\frac{P_{k}
\gamma(t,t_{y})}{N}\right)^{\frac{1}{\alpha}},
\end{equation}
and
\begin{equation}\label{eq26}
C_{\alpha}(\rho_{kH_{P}})\simeq\frac{N}{\alpha-1}\left(\frac{1-P_{k+1}}
{N}\right)^{\frac{1}{\alpha}},
\end{equation}
respectively for $t\ll N$ and $\alpha\in(1,2]$, where
$\gamma(t,t_{y})=\frac{(t-2t_{y})^{2}}{t}$ and
$P_{k}=\sin^{2}\theta_{k}$.

\noindent$\it{Proof}$. From Eq. (21) $|\psi_{kH_{O}}\rangle$ can be reexpressed as
  \begin{equation}\label{eq27}
   |\psi_{kH_{O}}\rangle=\frac{1}{\sqrt{N}}\left[\left(A_{k}\sqrt{N-t}-B_{k}\sqrt{t}\right)|0\rangle+
   \sum_{y\neq0}(t-2t_{y})\left(\frac{A_{k}}{\sqrt{N-t}}+\frac{B_{k}}{\sqrt{t}}\right)
   |y\rangle\right].
   \end{equation}
According to Eq. (22), we obtain
\begin{equation}\label{eq28}
|\psi_{kH_{P}}\rangle
\equiv H^{\otimes n}PH^{\otimes n} |\psi_{kO}\rangle
=|\psi_{k+1}\rangle
=A_{k+1}|\chi_{0}\rangle+B_{k+1}|\chi_{1}\rangle.
\end{equation}
By straightforward derivation we have
\begin{eqnarray}\label{eq29}
\rho_{kH_{O}}=|\psi_{kH_{O}}\rangle\langle\psi_{kH_{O}}|
=\frac{1}{N}\left[\left(A_{k}\sqrt{N-t}-B_{k}\sqrt{t}\right)^{2}|0\rangle\langle0|+\sum_{y\neq0}
(t-2t_{y})\right.
\nonumber\\
\times\left.\left(\frac{A_{k}}{\sqrt{N-t}}+\frac{B_{k}}{\sqrt{t}}\right)
\left(A_{k}\sqrt{N-t}-B_{k}\sqrt{t}\right)\left(|0\rangle\langle y|+|y\rangle\langle0|\right)\right.
\nonumber\\
\left.+\sum_{y,y^{'}\neq0}(t-2t_{y})(t-2t_{y^{'}})\left(\frac{A_{k}}
{\sqrt{N-t}}+\frac{B_{k}}{\sqrt{t}}\right)^{2}|y\rangle\langle y^{'}|\right],
\end{eqnarray}
  and $\rho_{kH_{P}}=|\psi_{kH_{P}}\rangle\langle\psi_{kH_{P}}|=\rho_{k+1}$.

Combining Eqs. (3) with (29) we have
  \begin{equation}\label{eq30}
  C_{\alpha}(\rho_{kH_{O}})=\frac{1}{\alpha-1}\left[\frac{\left(A_{k}\sqrt{N-t}-B_{k}
  \sqrt{t}\right)^{\frac{2}
  {\alpha}}}{N^{\frac{1}{\alpha}}}+(N-1)\frac{\left(t-2t_{y}\right)^{\frac{2}{\alpha}}
  \left(\frac{A_{k}}{\sqrt{N-t}}+
  \frac{B_{k}}{\sqrt{t}}\right)^{\frac{2}{\alpha}}}{N^{\frac{1}{\alpha}}}-1\right].
  \end{equation}
Since $\rho_{kH_{P}}=\rho_{k+1}$, we have $C_{\alpha}(\rho_{kH_{P}})= C_{\alpha}(\rho_{k+1})$ and
\begin{equation}\label{eq31}
C_{\alpha}(\rho_{kH_{P}})=\frac{1}{\alpha-1}\left[\left(1-P_{k+1}\right)^{\frac{1}{\alpha}}
(N-t)^{1-\frac{1}{\alpha}}+P_{k+1}^{\frac{1}{\alpha}}t^{1-\frac{1}{\alpha}}-1\right].
\end{equation}
Denote $\gamma(t,t_{y})=\frac{(t-2t_{y})^{2}}{t}$. Noting that $t\ll
N$, when $\alpha\in(0,1)$ we have (\ref{eq23}) and (\ref{eq24}), and we have (\ref{eq25}) and (\ref{eq26}) when
$\alpha\in(1,2]$. $\Box$

\noindent {\bf Remark 3.1} Set $\alpha=\frac{1}{2}$ in Eqs. (23) and
(24). The skew information of coherence of the states
$|\psi_{kH_{O}}\rangle$ and $|\psi_{kH_{P}}\rangle$ satisfy
\begin{equation}\label{eq32}
C_{s}(\rho_{kH_{O}})\simeq1-\frac{\left(\gamma(t,t_{y})P_{k}\right)^{2}}{N}
~~~\makebox{and}~~~C_{s}(\rho_{kH_{P}})\simeq1-\frac{P_{k+1}^{2}}{t},
\end{equation}
respectively. When $\alpha\rightarrow1$ in Eq. (\ref{eq30}),
according to Eq. (\ref{eq10}), the relative entropies of coherence
of the state $|\psi_{kH_{O}}\rangle$ and $|\psi_{kH_{P}}\rangle$
satisfy
\begin{equation}\label{eq33}
C_{r}(\rho_{kH_{O}}) \simeq\
\frac{\gamma(t,t_{y})P_{k}}{\ln2}\ln\frac{N}{\gamma(t,t_{y})P_{k}}
~~~\makebox{and}~~~C_{r}(\rho_{kH_{P}})\simeq
\frac{1-P_{k+1}}{\ln2}\ln\frac{N}{1-P_{k+1}},
\end{equation}
respectively. Based on Theorem 2, it is easy to see that the
coherence of the state $|\psi_{kH_{O}}\rangle$ is related to the
size of the database $N$, success probability $P_{k}$ and target
states. The coherence of the state $|\psi_{kH_{P}}\rangle$ is
dependent on $N$ and $P_{k}$.

\vskip0.1in

\noindent {\bf 4 Different target states}\\\hspace*{\fill}\\
In this section, we discuss the coherence of the state
$|\psi_{kH_{O}}\rangle$ for special target states, i.e., the
superposition state $|\chi_{1}\rangle$ is a product state, and the
target numbers are $t\leq4$. We have the following result.

\noindent {\bf Theorem 3} Suppose that the superposition state
$|\chi_{1}\rangle$ is a product state. Then when $t\ll N$ the Tsallis relative
$\alpha$ entropy of coherence of the state
$|\psi_{kH_{O}}\rangle$ is given by
\begin{equation}\label{eq34}
C_{\alpha}(\rho_{kH_{O}})\simeq\frac{1}{\alpha-1}
\left[\left(\frac{P_{k}}{t}\right)^{\frac{1}{\alpha}}N^{1-\frac{1}{\alpha}}-1\right],
\end{equation}
for $\alpha\in(0,1)$, and
\begin{equation}\label{eq35}
C_{\alpha}(\rho_{kH_{O}})\simeq\frac{N}{\alpha-1}\left(\frac{P_{k}}{Nt}\right)^
{\frac{1}{\alpha}},
\end{equation}
for $\alpha\in(1,2]$.

\noindent $\it{Proof}$. Denote
$|\chi_{0H}\rangle=H^{\otimes n}|\chi_{0}\rangle$ and
$|\chi_{1H}\rangle=H^{\otimes n}|\chi_{1}\rangle$. Since
$|\chi_{1}\rangle=\frac{1}{\sqrt{t}}\sum_{x_{s}}|x_{s}\rangle$ is the
product of $n$ single-qubit states of the forms either $|0\rangle$, $|1\rangle$ or
$(|0\rangle\pm|1\rangle)/\sqrt{2}$, $|\psi_{kH_{O}}\rangle$ can be written as
\begin{align}\label{eq36}
|\psi_{kH_{O}}\rangle\notag
=&H^{\otimes n}(A_{k}|\chi_{0}\rangle-B_{k}|\chi_{1}\rangle)\\
=&H^{\otimes n}A_{k}|\chi_{0}\rangle-H^{\otimes n}B_{k}|\chi_{1}\rangle\notag\\
=&A_{k}|\chi_{0H}\rangle-B_{k}|\chi_{1H}\rangle,
\end{align}
and
\begin{align}\label{eq37}
  \rho_{kH_{O}}=A_{k}^{2}|\chi_{0H}\rangle\langle \chi_{0H}|-A_{k}B_{k}(|\chi_{1H}
  \rangle\langle \chi_{0H}|\notag\\
  +|\chi_{0H}\rangle\langle\chi_{1H}|)+B_{k}^{2}|\chi_{1H}\rangle\langle \chi_{1H}|,
  \end{align}
where
\begin{equation*}\label{eq}
|\chi_{0H}\rangle =\sum_{y=0}^{N-1}\sum_{x\in
\{x_{n}\}}(-1)^{xy}\frac{1}{\sqrt{N(N-t)}}|y\rangle,
\end{equation*}
and
\begin{equation*}\label{eq}
|\chi_{1H}\rangle =\sum_{y=0}^{N-1}\sum_{x\in
\{x_{s}\}}(-1)^{xy}\frac{1} {\sqrt{Nt}}|y\rangle.
\end{equation*}

After a simple transformation, we have
\begin{equation}\label{eq38}
|\chi_{0H}\rangle
=\sqrt{\frac{N-t}{N}}|0\rangle+\sum_{y\neq0}\frac{t-2t_{y}}{\sqrt{N(N-t)}}|y\rangle,
\end{equation}
and
\begin{equation}\label{eq39}
|\chi_{1H}\rangle
=\sqrt{\frac{t}{N}}|0\rangle+\sum_{y\neq0}\frac{2t_{y}-t}{\sqrt{Nt}}|y\rangle.
\end{equation}
Consequently, we derive that
\begin{eqnarray}\label{eq40}
&&C_{\alpha}(\rho_{kH_{O}})
=\frac{1}{\alpha-1}\left[\left(\frac{A_{k}^{2}(N-t)}{N}-2A_{k}
B_{k}\sqrt{\frac{(N-t)t}{N^{2}}}+\frac{B_{k}^{2}t}{N}\right)^{\frac{1}{\alpha}}
\right.
\nonumber\\
&&\left.+(N-1)\left(A_{k}^{2}\frac{(t-2t_{y})^{2}}{N(N-t)}-2A_{k}B_{k}\frac{(t-2t_{y})
(2t_{y}-t)}{N\sqrt{(N-t)t}}+B_{k}^{2}\frac{(2t_{y}-t)^{2}}{Nt}\right)^{\frac{1}{\alpha}}-1\right].
\end{eqnarray}
When $t\ll N$, $\alpha\in(0,1)$, and $P_{k}=\sin^{2}\theta_{k}$, we have
\begin{align}\label{eq41}
C_{\alpha}(\rho_{kH_{O}})\notag
\simeq&\frac{1}{\alpha-1}\left[\left(\frac{B_{k}^{2}t}{N}\right)^{\frac{1}
{\alpha}}+\left(N-1\right)\left(B_{k}^{2}\frac{(2t_{y}-t)^{2}}{Nt}\right)^{\frac{1}{\alpha}}-1\right]\\
\simeq&\frac{1}{\alpha-1}\left[\left(\frac{P_{k}}{t}\right)^{\frac{1}{\alpha}}N^{1-\frac{1}{\alpha}}-1\right],
\end{align}
and when $t\ll N$, $\alpha\in(1,2]$, $C_{\alpha}(\rho_{kH_{O}})$ takes
the form
\begin{align}\label{eq42}
C_{\alpha}(\rho_{kH_{O}})\notag
\simeq&\frac{1}{\alpha-1}\left[\left(\frac{B_{k}^{2}t}{N}\right)^{\frac{1}
{\alpha}}+(N-1)\left(B_{k}^{2}\frac{(2t_{y}-t)^{2}}{Nt}\right)^{\frac{1}{\alpha}}-1\right]\\
\simeq&\frac{N}{\alpha-1}\left(\frac{P_{k}}{Nt}\right)^{\frac{1}{\alpha}}.
\end{align}
This completes the proof. $\Box$

In addition, for the case that $|\chi_{1}\rangle$ is a product state, $t\ll N$ and $\alpha=\frac{1}{2}$ in Eq. (\ref{eq34}), the skew information of coherence
of the state $|\psi_{kH_{O}}\rangle$ is given by
\begin{equation}\label{eq43}
C_{s}(\rho_{kH_{O}}) \simeq1-\frac{P_{k}^{2}}{t^{2}N}.
\end{equation}
Taking the limit $\alpha\rightarrow1$ in Eq. (\ref{eq40}), the relative
entropy of coherence of the state $|\psi_{kH_{O}}\rangle$ is given
by
\begin{align}\label{eq44}
C_{r}(\rho_{kH_{O}})\notag
\simeq&-\frac{P_{k}t}{N\ln2}\ln\frac{P_{k}t}{N}-\frac{P_{k}}{t\ln2}\ln\frac{P_{k}}{Nt}\\
\simeq&\frac{P_{k}}{t\ln2}\ln\frac{Nt}{P_{k}}.
\end{align}

We now explore the coherence of the state $|\psi_{kH_{O}}\rangle$ when
the number of the target states is small ($t\leq4$).\\
$\bullet$ When $t$=1, the database has one single target state,
$|\chi_{1}\rangle$ is always a product state. According to Theorem 3 we get
\begin{equation}\label{eq45}
   C_{\alpha}(\rho_{kH_{O}})\simeq
\frac{1}{\alpha-1}\left[P_{k}^{\frac{1}{\alpha}}N^{1-\frac{1}{\alpha}}-1\right],
   \end{equation}
when $\alpha\in(0,1)$, and
\begin{equation}\label{eq46}
   C_{\alpha}(\rho_{kH_{O}})\simeq\frac{N}{\alpha-1}\left(\frac{P_{k}}{N}\right)
^{\frac{1}{\alpha}},
   \end{equation}
when $\alpha\in(1,2]$.\\
$\bullet$ When $t$=2, we denote the two target states by
$|x_{11}\rangle$ and $|x_{12}\rangle$, respectively. We have $|t-2t_{y}|=1$
for either $|x_{11}\rangle =0$ or $|x_{11}\rangle\neq0$ \cite{PMQ}. Thus
$\gamma(t,t_{y})=\frac{1}{2}$.\\
(1) When $\alpha\in(0,1)$, according to Eq. (\ref{eq23}) we have
   \begin{equation}\label{eq47}
   C_{\alpha}(\rho_{kH_{O}})\simeq\frac{1}{\alpha-1}\left[\left(\frac{P_{k}}{2}\right)^{\frac{1}
   {\alpha}}N^{1-\frac{1}{\alpha}}-1\right].
   \end{equation}
(2) When $\alpha\in(1,2]$, according to Eq. (\ref{eq25}) we obtain
   \begin{equation}\label{eq48}
   C_{\alpha}(\rho_{kH_{O}})\simeq\frac{N}{\alpha-1}\left(\frac{P_{k}}{2N}\right)
   ^{\frac{1}{\alpha}}.
   \end{equation}
$\bullet$ When $t$=3, $|t-2t_{y}|=\frac{3}{2}$ and $\gamma(t,t_{y})=\frac{3}{4}$.\\
(1) According to Eq. (\ref{eq23}) we get
   \begin{equation}\label{eq49}
   C_{\alpha}(\rho_{kH_{O}})\simeq\frac{1}{\alpha-1}\left[\left(\frac{3P_{k}}{4}\right)^{\frac{1}
   {\alpha}}N^{1-\frac{1}{\alpha}}-1\right],
   \end{equation}
for $\alpha\in(0,1)$,\\
(2) for $\alpha\in(1,2]$,
   \begin{equation}\label{eq50}
   C_{\alpha}(\rho_{kH_{O}})\simeq\frac{N}{\alpha-1}\left(\frac{3P_{k}}{4N}\right)
   ^{\frac{1}{\alpha}}.
   \end{equation}
$\bullet$ When $t$=4, there are two cases that $|\chi_{1}\rangle$ is either a product state or not.\\
Case 1: $|\chi_{1}\rangle$ is a product state. It is observed that $|t-2t_{y}|=1$ \cite{PMQ} and $\gamma(t,t_{y})=\frac{1}{4}$.\\
(1) When $\alpha\in(0,1)$ we obtain
   \begin{equation}\label{eq51}
   C_{\alpha}(\rho_{kH_{O}})\simeq\frac{1}{\alpha-1}\left[\left(\frac{P_{k}}{4}\right)^{\frac{1}
   {\alpha}}N^{1-\frac{1}{\alpha}}-1\right].
   \end{equation}
(2) When $\alpha\in(1,2]$ the coherence of the state $|\psi_{kH_{O}}\rangle$ is
   \begin{equation}\label{eq52}
   C_{\alpha}(\rho_{kH_{O}})\simeq\frac{N}{\alpha-1}\left(\frac{P_{k}}{4N}\right)
   ^{\frac{1}{\alpha}}.
   \end{equation}
Case 2: $|\chi_{1}\rangle$ is not a product state. It is observed that $|t-2t_{y}|=\frac{3}{2}$ \cite{PMQ} and $\gamma(t,t_{y})=\frac{9}{16}$.\\
(1) When $\alpha\in(0,1)$ we have
   \begin{equation}\label{eq53}
   C_{\alpha}(\rho_{kH_{O}})\simeq\frac{1}{\alpha-1}\left[\left(\frac{9P_{k}}{16}\right)^{\frac{1}
   {\alpha}}N^{1-\frac{1}{\alpha}}-1\right].
   \end{equation}
(2) When $\alpha\in(1,2]$, the coherence of the state $|\psi_{kH_{O}}\rangle$ is
   \begin{equation}\label{eq54}
   C_{\alpha}(\rho_{kH_{O}})\simeq\frac{N}{\alpha-1}\left(\frac{9P_{k}}{16N}\right)
   ^{\frac{1}{\alpha}}.
   \end{equation}

\noindent {\bf Remark 4.1} (1) When $t=1$ and $\alpha=\frac{1}{2}$ in Eq.
(\ref{eq45}), according to Eq. (\ref{eq44}) the skew information of coherence
and the relative entropy of coherence of the state
$|\psi_{kH_{O}}\rangle$ are
\begin{equation*}\label{eq}
C_{s}(\rho_{kH_{O}})\simeq1-\frac{P_{k}^{2}}{N}~~~\makebox{and}~~~
C_{r}(\rho_{kH_{O}})\simeq\frac{P_{k}}{\ln2}\ln\frac{N}{P_{k}},
\end{equation*}
respectively. \\
(2) When $t=2$, $\alpha=\frac{1}{2}$ in Eq.
(\ref{eq47}), according to Eq. (\ref{eq33}) the skew information of
coherence and the relative entropy of coherence of the state
$|\psi_{kH_{O}}\rangle$ are
\begin{equation*}\label{eq}
C_{s}(\rho_{kH_{O}})\simeq1-\frac{P_{k}^{2}}{4N}~~~\makebox{and}~~~
C_{r}(\rho_{kH_{O}})\simeq\frac{P_{k}}{2\ln2}\ln\frac{2N}{P_{k}},
\end{equation*}
respectively. \\
(3) When $t=3$ and $\alpha=\frac{1}{2}$ in Eq. (\ref{eq49}),
according to Eq. (\ref{eq33}) the skew information of coherence and
the relative entropy of coherence of the state
$|\psi_{kH_{O}}\rangle$ are
\begin{equation*}\label{eq}
C_{s}(\rho_{kH_{O}})\simeq1-\frac{9P_{k}^{2}}{16N}~~~\makebox{and}~~~
C_{r}(\rho_{kH_{O}})\simeq\frac{3P_{k}}{4\ln2}\ln\frac{4N}{3P_{k}},
\end{equation*}
respectively. \\
(4) When $t=4$, there are two cases.\\
Case 1: Set $\alpha=\frac{1}{2}$ in Eq. (\ref{eq51}). According to
Eq. (\ref{eq33}) the skew information of coherence and the relative entropy
of coherence of the state $|\psi_{kH_{O}}\rangle$ are
\begin{equation*}\label{eq}
C_{s}(\rho_{kH_{O}})\simeq1-\frac{P_{k}^{2}}{16N}~~~\makebox{and}~~~
C_{r}(\rho_{kH_{O}})\simeq\frac{P_{k}}{4\ln2}\ln\frac{4N}{P_{k}},
\end{equation*}
respectively. \\
Case 2: Set $\alpha=\frac{1}{2}$ in Eq. (\ref{eq53}). According to
Eq. (\ref{eq33}), the skew information of coherence and the relative
entropy of coherence of the state $|\psi_{kH_{O}}\rangle$ are
\begin{equation*}\label{eq}
C_{s}(\rho_{kH_{O}})\simeq1-\frac{81P_{k}^{2}}{256N}~~~\makebox{and}~~~
C_{r}(\rho_{kH_{O}})\simeq\frac{9P_{k}}{16\ln2}\ln\frac{16N}{9P_{k}},
\end{equation*}
respectively.

According to Theorem 3 and Eqs. (\ref{eq51}), (\ref{eq52}), (\ref{eq53}) and (\ref{eq54}), when
$\alpha\in(1,2]$ the Tsallis relative $\alpha$ entropy of coherence of
the state $|\psi_{kH_{O}}\rangle$ is larger when the superposition
state of the targets is an entangled one.
However, when $\alpha\in(0,1)$ it is found that the coherence is
smaller when the superposition state of the targets is entangled.

\noindent {\bf Conjecture} The Tsallis relative $\alpha$ entropy of
coherence of the state $|\psi_{kH_{O}}\rangle$ relies on the size of
the database $N$, the success probability and the target states.\\
(1) For $\alpha\in(0,1)$, the coherence of $|\psi_{kH_{O}}\rangle$
reaches the lower bound when $t=1$, and coherence of
$|\psi_{kH_{O}}\rangle$ reaches the upper bound when $|\chi_{1}\rangle$
is a product state. It holds that
\begin{equation*}\label{eq}
\frac{1}{\alpha-1}\left[P_{k}^{\frac{1}{\alpha}}
N^{1-\frac{1}{\alpha}}-1\right]\leq C_{\alpha}
(\rho_{kH_{O}})
\leq\frac{1}{\alpha-1}
\left[\left(\frac{P_{k}}{t}\right)^{\frac{1}{\alpha}}N^{1-\frac{1}{\alpha}}-1\right].
\end{equation*}
(2) For $\alpha\in(1,2]$, the coherence of $|\psi_{kH_{O}}\rangle$
reaches the upper bound when $t=1$, and coherence of
$|\psi_{kH_{O}}\rangle$  reaches the lower bound when
$|\chi_{1}\rangle$ is a product state. It holds that
\begin{equation*}\label{eq}
\frac{N}{\alpha-1}\left(\frac{P_{k}}{Nt}\right)^{\frac{1}{\alpha}}\leq C_{\alpha}
(\rho_{kH_{O}})
\leq\frac{N}{\alpha-1}\left(\frac{P_{k}}{N}\right)^{\frac{1}{\alpha}}.
\end{equation*}

\vskip0.1in

\noindent {\bf 5 Production and depletion of Tsallis relative $\alpha$ entropy of coherence }\\\hspace*{\fill}\\
The coherence of a state changes when the operators $H_{P}$ and $H_{O}$ are applied. In this section, we investigate how the coherence changes before and after these operators are applied, i.e., the production and depletion of the Tsallis relative $\alpha$ entropy of
coherence for $\alpha\in(1,2]$. In order to clarify the variations
and connections of operator coherence in GSA, we also provide some examples and diagrammatic sketches related to the coherence dynamics.

We first introduce the following definitions. The variations of operator coherence between two consecutive iterations of $H_{O}$, $H_{P}$ and $G$ in GSA are defined as
\begin{equation}\label{eq55}
\Delta C^{\alpha}(\rho_{kG})\equiv C_{\alpha}^{\alpha}(\rho_{k+1})-C_{\alpha}^{\alpha}(\rho_{k}),
\end{equation}
\begin{equation}\label{eq56}
\Delta C^{\alpha}(\rho_{kH_{O}})\equiv C_{\alpha}^{\alpha}(\rho_{(k+1)H_{O}})-C_{\alpha}
^{\alpha}(\rho_{kH_{O}}),
\end{equation}
\begin{equation}\label{eq57}
\Delta C^{\alpha}(\rho_{kH_{P}})\equiv C_{\alpha}^{\alpha}(\rho_{(k+1)H_{P}})-C_{\alpha}
^{\alpha}(\rho_{kH_{P}}).
\end{equation}
The variations of suboperator coherence of each basic operator $H_{O}$ and $H_{P}$ in one Grover iteration are defined as
\begin{equation}\label{eq58}
\Delta C^{\alpha}(\rho_{k\Delta H_{O}})\equiv C_{\alpha}^{\alpha}(\rho_{kH_{O}})-C_{\alpha}
^{\alpha}(\rho_{k}),
\end{equation}
\begin{equation}\label{eq59}
\Delta C^{\alpha}(\rho_{k\Delta H_{P}})\equiv C_{\alpha}^{\alpha}(\rho_{kH_{P}})-C_{\alpha}
^{\alpha}(\rho_{kH_{O}}).
\end{equation}

Concerning the production and depletion of coherence for basic operators in GSA, we have the following conclusion.

\noindent {\bf Theorem 4} For $\alpha\in(1,2]$ and $t\ll N$, the variations
and connections of operator coherence between two consecutive
iterations of $H_{O}$, $H_{P}$ and $G$ in GSA are given by
\begin{equation*}\label{eq}
\Delta C^{\alpha}(\rho_{kG})\simeq \frac{N^{\alpha-1}}{(\alpha-1)^{\alpha}}(P_{k}-P_{k+1})<0,
\end{equation*}
\begin{equation*}\label{eq}
\Delta C^{\alpha}(\rho_{kH_{O}})\simeq\frac{N^{\alpha-1}}{(\alpha-1)^{\alpha}}
\gamma(t,t_{y})(P_{k+1}-P_{k})>0,
\end{equation*}
\begin{equation}\label{eq60}
\Delta C^{\alpha}(\rho_{kH_{P}})\simeq \frac{N^{\alpha-1}}{(\alpha-1)^{\alpha}}(P_{k+1}-P_{k+2})<0,
\end{equation}
and
\begin{equation}\label{eq61}
\Delta C^{\alpha}(\rho_{kG})\simeq\Delta C^{\alpha}(\rho_{(k-1)H_{P}})\simeq-\frac{1}{\gamma(t,t_{y})}
\Delta C^{\alpha}(\rho_{kH_{O}}).
\end{equation}\\
$\it{Proof}$. Combining Eqs. (\ref{eq8}) with (\ref{eq55}), it is easy to obtain
\begin{equation*}\label{eq}
\Delta C^{\alpha}(\rho_{kG})\simeq \frac{N^{\alpha-1}}{(\alpha-1)^{\alpha}}(P_{k}-P_{k+1}).
\end{equation*}
Since $P_{k}<P_{k+1}$, we have
\begin{equation*}\label{eq}
\Delta C^{\alpha}(\rho_{kG})\simeq \frac{N^{\alpha-1}}{(\alpha-1)^{\alpha}}(P_{k}-P_{k+1})<0.
\end{equation*}
Similarly, according to Eqs. (\ref{eq25}) and (\ref{eq56}), it is easy to get
\begin{equation*}\label{eq}
\Delta
C^{\alpha}(\rho_{kH_{O}})\simeq\frac{N^{\alpha-1}}{(\alpha-1)^{\alpha}}
\gamma(t,t_{y})(P_{k+1}-P_{k})>0.
\end{equation*}
Combining Eqs. (\ref{eq26}) with (\ref{eq57}), we have
\begin{equation*}\label{eq}
\Delta C^{\alpha}(\rho_{kH_{P}})\simeq \frac{N^{\alpha-1}}{(\alpha-1)^{\alpha}}
(P_{k+1}-P_{k+2})<0.
\end{equation*}
Obviously, we have the following relationship among these variations,
\begin{equation*}\label{eq}
\Delta C^{\alpha}(\rho_{kG})\simeq\Delta C^{\alpha}(\rho_{(k-1)H_{P}})\simeq-\frac{1}
{\gamma(t,t_{y})} \Delta C^{\alpha}(\rho_{kH_{O}}).
\end{equation*}
$\Box$

According to Theorem 4, the operator coherence between two
consecutive iterations of $H_{P}$ and $G$ is depleted. Both of them rely on
the size of the database $N$ and the success probability.
Correspondingly, the operator coherence between two consecutive
iterations of $H_{O}$ is produced, which relies on the
size of the database $N$, the success probability and target states.

\noindent {\bf Theorem 5} For $\alpha\in(1,2]$ and $t\ll N$, the functions
$\Delta C^{\alpha}(\rho_{kH_{P}})$ and $\Delta C^{\alpha}(\rho_{kH_{O}})$
have a turning point. The variations of the
suboperator coherence of each basic operator $H_{O}$ and $H_{P}$ in
one Grover iteration are given by
\begin{equation}\label{eq62}
\Delta C^{\alpha}(\rho_{k\Delta H_{O}})\simeq \frac{N^{\alpha-1}}{(\alpha-1)^{\alpha}}
\left[\left(\gamma(t,t_{y})+1\right)P_{k}-1\right],
\end{equation}
\begin{equation}\label{eq63}
\Delta C^{\alpha}(\rho_{k\Delta H_{P}})\simeq \frac{N^{\alpha-1}}{(\alpha-1)^{\alpha}}
\left[1-(\gamma(t,t_{y})P_{k}+P_{k+1})\right].
\end{equation}

\noindent $\it{Proof}$. According to Eqs. (\ref{eq8}), (\ref{eq25}) and (\ref{eq58}), we have
(\ref{eq62}). Similarly, substituting Eqs. (\ref{eq25}) and (\ref{eq26}) into (\ref{eq59}), we obtain (\ref{eq63}). There exists a turning point $k_{T}$ at which $\Delta
C^{\alpha}(\rho_{k\Delta H_{O}})=0$. This is equivalent to
$(\gamma(t,t_{y})+1)P_{k}-1=0$, namely,
\begin{equation*}\label{eq}
k_{T}\simeq\left[\frac{\arcsin\sqrt{\frac{1}{\gamma(t,t_{y})+1}}}{2\theta}\right].
\end{equation*}
Similar results can be obtained for $\Delta C^{\alpha}(\rho_{k\Delta H_{P}})=0$. $\Box$

According to Theorem 5, for $t\ll N$ we have $\Delta
C^{\alpha}(\rho_{k\Delta H_{O}})<0$ and $\Delta
C^{\alpha}(\rho_{k\Delta H_{P}})>0$ when
$(\gamma(t,t_{y})+1)P_{k}<1$, and $\Delta C^{\alpha}(\rho_{k\Delta
H_{O}})>0$ and $\Delta C^{\alpha}(\rho_{k\Delta H_{P}})<0$ when
$(\gamma(t,t_{y})+1)P_{k}>1$. The Tsallis relative $\alpha$ entropy
of coherence of $H_{O}$ and $H_{P}$ show different effects that one
depletes coherence and the other produces coherence. Moreover, the
operator coherence of $H^{\otimes n}$ is not always produced or
depleted, but depleted and produced alternatively. Before the
turning point, the operator coherence of $H_{O}$ is depleting, and
the operator coherence of $H_{P}$ is producing. However, the
situation is reversed after the turning point.

We now use examples and plots to illustrate the
characters of coherence of the state after each basic operator is
applied in GSA, how these operators contribute to coherence, and the
relationships among the coherence of the operators and the success
probability.

\noindent {\bf Example 1} Suppose that the qubit numbers are $n$=16
and the target numbers are $t$=2. In this case
$\gamma(t,t_{y})=\frac{1}{2}$. Based on Theorem 5, the
suboperator coherence of $O$ and $P$ in one Grover iteration are
unaltered. When $k< k_{T}$ the suboperator coherence of $H_{O}$ in
one Grover iteration is depleting, and the suboperator coherence of
$H_{P}$ is producing. However, when $k> k_{T}$ the situation is
reversed. Fig. 1 shows the variations of the suboperator coherence of
each basic operator $O$, $H_{O}$, $P$ and $H_{P}$ in one Grover
iteration. For clarity in Fig. 1 we use
$\frac{((\alpha-1)C_{\alpha})^{\alpha}}{N^{\alpha-1}}$ as the vertical axis.
From Fig. 1 we see that before the
turning point, the suboperator coherence of each basic operator $O$,
$H_{O}$, $P$ and $H_{P}$ in one Grover iteration are unchanged,
decreased, unchanged and increased, respectively; while after the turning
point, they are unchanged, increased, unchanged and decreased, respectively.
\begin{figure}[ht]\centering
\subfigure[] {\begin{minipage}[figure1a]{0.3\linewidth}
\includegraphics[width=1.0\textwidth]{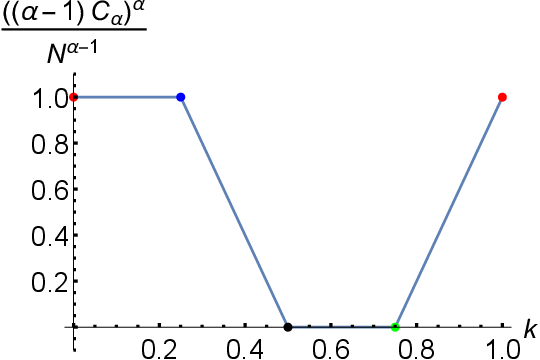}
\end{minipage}}
\subfigure[] {\begin{minipage}[figure1b]{0.3\linewidth}
\includegraphics[width=1.0\textwidth]{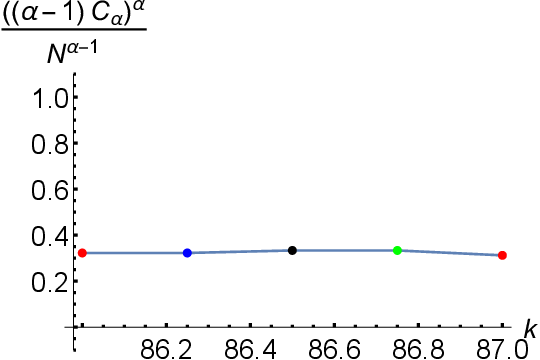}
\end{minipage}}
\subfigure[] {\begin{minipage}[figure1c]{0.3\linewidth}
\includegraphics[width=1.0\textwidth]{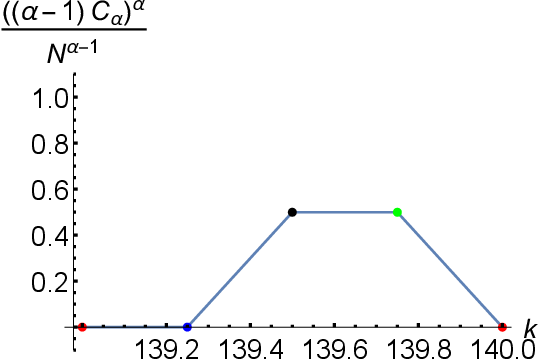}
\end{minipage}}
\caption{{The coherence dynamics in one Grover iteration. The red, blue,
black and green dots are the coherences of $O$, $H_{O}$, $P$ and $H_{P}$, respectively.
The variations of the suboperator coherence before the
turning point ($\mathbf{a}$), at the turning point ($\mathbf{b}$) and after the turning point ($\mathbf{c}$). \label{fig:Fig1}}}
\end{figure}

Fig. 2 shows that the success probability $P_{k}$
increases with the increase of the number of iterations. According
to Eqs. (\ref{eq25}) and (\ref{eq26}), the operator coherence of $H^{\otimes n}$
vibrates between $1-P_{k+1}$ and $\gamma(t,t_{y})P_{k}$. The relations between
the operator coherence of $H^{\otimes n}$ and the operator coherence of $H _{O}$ and $H _{P}$ are shown in Fig. 3. The intersection of the two lines is the turning point.
\begin{figure}[ht]
\centering
{\begin{minipage}[figure2]{0.5\linewidth}
\includegraphics[width=0.95\textwidth]{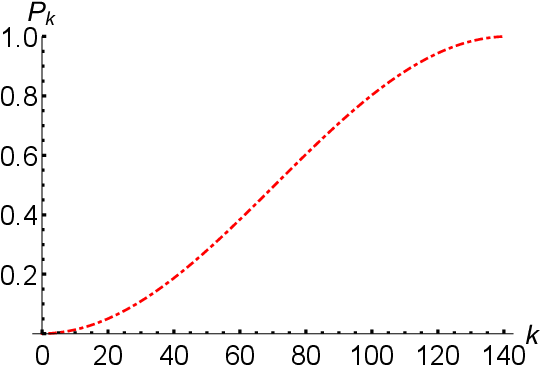}
\end{minipage}}
\caption{{The variations of success probability $P_{k}$ (red dot-dashed line) as a function of the number of iterations $k$. \label{fig:Fig2}}}
\end{figure}
\begin{figure}[ht]
\centering
{\begin{minipage}[figure3]{0.5\linewidth}
\includegraphics[width=0.95\textwidth]{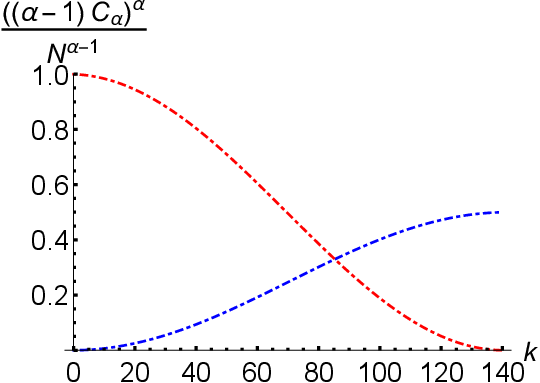}
\end{minipage}}
\caption{{The operator coherence of $H^{\otimes n}$. The blue
dot-dashed line and red dot-dashed line represent the operator
coherence of $H _{O}$ and $H _{P}$, respectively. \label{fig:Fig3}}}
\end{figure}

In Fig. 4, we illustrate the relationships of $\Delta C^{\alpha}
(\rho_{kG})$, $\Delta C^{\alpha}(\rho_{kH_{P}})$ and $\Delta
C^{\alpha}(\rho_{kH_{O}})$ between two consecutive iterations.
\begin{figure}[ht]
\centering
{\begin{minipage}[figure4]{0.5\linewidth}
\includegraphics[width=0.95\textwidth]{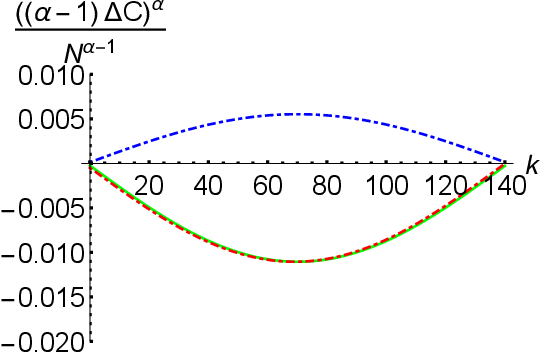}
\end{minipage}}
\caption{{The operator coherence of $G$ (green), $H_{P}$ (red dot-dashed) and $H_{O}$ (blue dot-dashed) between two consecutive iterations.  \label{fig:Fig4}}}
\end{figure}

According to Theorem 3, we have $\Delta
C^{\alpha}(\rho_{kG})\simeq\Delta C^{\alpha}
(\rho_{(k-1)H_{P}})\simeq-\frac{1}{\gamma(t,t_{y})} \Delta
C^{\alpha}(\rho_{kH_{O}})$, which are zero at the beginning and the end. The
connections of the suboperator coherence of $ H_{P}$ and $H_{O}$ in one
Grover iteration are shown in Fig. 5. Moreover, the coherence of
$H_{P}$ ($H_{O}$) is 1 ($-1$) at the beginning and $-\frac{1}{2}$ ($\frac{1}{2}$) at the end.
\begin{figure}[ht]
\centering
{\begin{minipage}[figure5]{0.5\linewidth}
\includegraphics[width=0.95\textwidth]{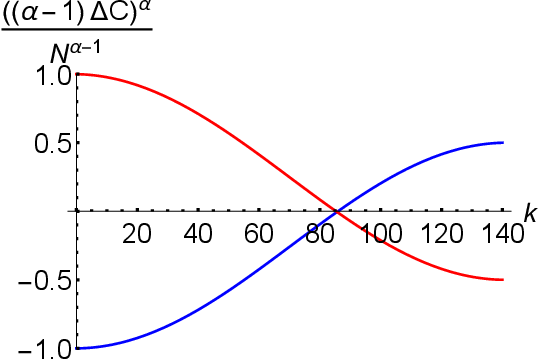}
\end{minipage}}
\caption{{The suboperator coherence of $ H_{P}$ (red) and $H_{O}$ (blue) in one Grover iteration.
\label{fig:Fig5}}}
\end{figure}

\noindent {\bf Example 2} Consider that the qubit numbers are $n$=18
and the target numbers are $t$=4. In this case
$\gamma(t,t_{y})=\frac{9}{16}$ when the superposition state of
targets is entangled and $\gamma(t,t_{y})=\frac{1}{4}$ when
superposition state is a product one. From Theorem 3
the Tsallis relative
$\alpha$ entropy of coherence of $|\psi_{kH_{O}}\rangle$ is larger
when the superposition state of targets is entangled for $\alpha\in(1,2]$.
When the superposition state of targets is
entangled, we show in Fig. 6 the relationships between the
operator coherence of $H_{P}$ and $H_{O}$, the relationships among
$\Delta C^{\alpha}(\rho_{kG})$, $\Delta C^{\alpha}(\rho_{kH_{P}})$
and $\Delta C^{\alpha}(\rho_{kH_{O}})$ between two consecutive
iterations, and the connections between the suboperator coherence of $
H_{P}$ and $H_{O}$ in one Grover iteration. For comparison, we also
present the corresponding results when the superposition state of targets is a
product one.

Note that Tsallis relative $\alpha$ entropy of
coherence incorporates two important coherence quantifiers, the
relative entropy of coherence and the skew information of coherence,
so using such technical methods of Tsallis relative $\alpha$ entropy
to study coherent dynamics may yield more information about the
properties of different coherent measures in GSA. It can also be
seen from the examples that entanglement has an important
contribution to operator coherence in GSA.
\begin{figure}[ht]\centering
\subfigure[] {\begin{minipage}[figure6a]{0.3\linewidth}
\includegraphics[width=1.0\textwidth]{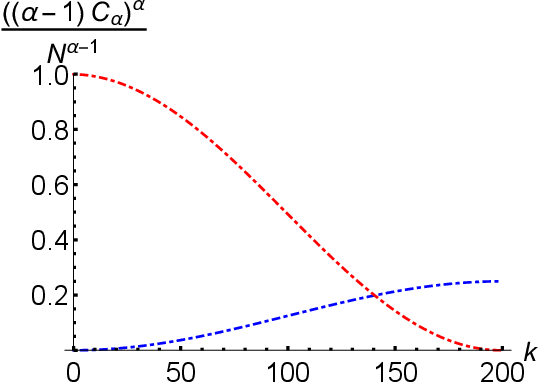}
\end{minipage}}
\subfigure[] {\begin{minipage}[figure6b]{0.3\linewidth}
\includegraphics[width=1.0\textwidth]{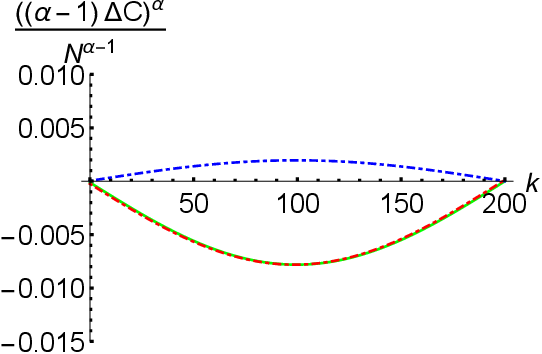}
\end{minipage}}
\subfigure[] {\begin{minipage}[figure6c]{0.3\linewidth}
\includegraphics[width=1.0\textwidth]{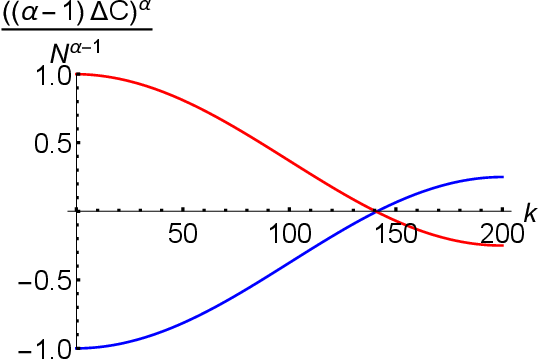}
\end{minipage}}
\subfigure[] {\begin{minipage}[figure6d]{0.3\linewidth}
\includegraphics[width=1.0\textwidth]{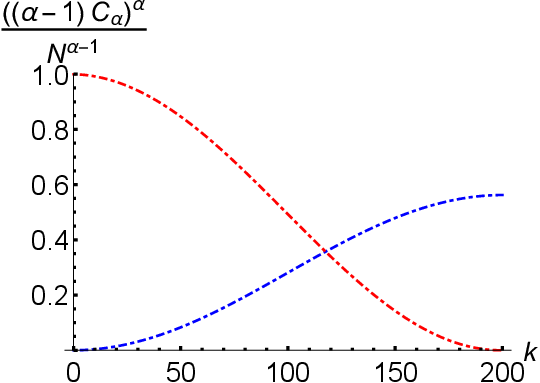}
\end{minipage}}
\subfigure[] {\begin{minipage}[figure6e]{0.3\linewidth}
\includegraphics[width=1.0\textwidth]{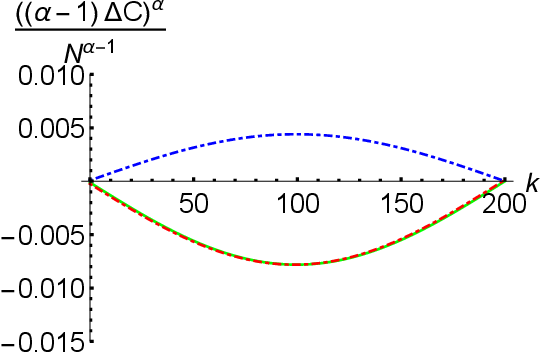}
\end{minipage}}
\subfigure[] {\begin{minipage}[figure6f]{0.3\linewidth}
\includegraphics[width=1.0\textwidth]{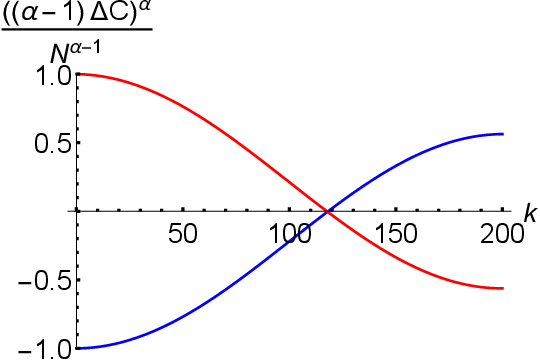}
\end{minipage}}
\caption{{Subfigures $\mathbf{a}$, $\mathbf{b}$ and $\mathbf{c}$ ($\mathbf{d}$, $\mathbf{e}$ and $\mathbf{f}$) are for the case that the superposition state of targets
is a product one (an entangled one). ($\mathbf{a}$, $\mathbf{d}$) The relationships of the
operator coherence of $H_{P}$ and $H_{O}$. ($\mathbf{b}$, $\mathbf{e}$) The
relationships of $\Delta C^{\alpha}(\rho_{kG})$, $\Delta C^{\alpha}(\rho_{kH_{P}})$ and
$\Delta C^{\alpha}(\rho_{kH_{O}})$ between two consecutive iterations.
($\mathbf{c}$, $\mathbf{f}$) The connections of the suboperator coherence of
$ H_{P}$ and $H_{O}$ in one Grover iteration.
\label{fig:Fig6}}}
\end{figure}

\vskip0.1in

\noindent {\bf 6 Comparison with previous works}\\\hspace*{\fill}\\
In order to clarify the contribution of this paper,
we compare our work with previous related works in this section.

In \cite{PMQ}, the authors have investigated the
$l_{1}$ norm of coherence of the states after each of $O$, $P$ and
$H^{\otimes n}$ is applied in one $G$ iteration, and discussed the
number of coherence for different cases of the target states. It is
shown that the coherence is monotone decreasing with the increase of
the success probability, and it is proved that the coherence is
vibrating, the overall effect is that coherence is in depletion.
Moreover, the coherence is larger when the superposition state of
targets is an entangled one.

In this work, we study coherence dynamics in GSA
based on Tsallis relative $\alpha$ entropy. The amount of coherence
of first $H^{\otimes n}$ depends on the size of the database $N$,
success probability and target states, and the coherence of two
$H^{\otimes n}$ have different effects that one depletes coherence
and the other produces coherence. Coherence is not always produced
or depleted, but depleted and produced in turn. When
$\alpha\in(0,1)$, the coherence is smaller when the superposition
state of targets is an entangled one, and the coherence reduction is
not monotonic, but related to the parameter $\alpha$ with the
increase of the success probability, which are different from
\cite{PMQ}.

On the other hand, in \cite{MH}, the Grover
iteration has been decomposed into two basic operators $R$ and $O$,
where $R=H^{\otimes n}PH^{\otimes n}$. It is demonstrated that
$H^{\otimes n}$ does not change entanglement, and there exists a
turning point during the application of the algorithm. Before the
turning point, the entanglement always increases when the operator
$O$ is applied, and the effect of the operator $R$ can be almost
ignored on the level of entanglement. After the turning point, the
$R$ remarkably decreases entanglement, and $O$ increases
entanglement. In our work, we study the coherence of the states on
essential operator level, and show the operators $O$ and $P$ do not
change the coherence. In addition, we also obtain a turning point
when $\alpha\in(1,2]$. Before the turning point, the operator
coherence of $H_{O}$ is depleting, and the operator coherence of
$H_{P}$ is producing. After the turning point, the situation is
reversed.

\vskip0.1in

\noindent {\bf 7 Conclusions and
discussions}\\\hspace*{\fill}\\
We have explored the coherence dynamics in Grover's search algorithm
(GSA) based on Tsallis relative $\alpha$ entropy for
$\alpha\in(0,1)\cup(1,2]$, and proved that the coherence decreases
with the increase of the success probability. We have derived the
complementarity relations between Tsallis relative $\alpha$ entropy
of coherence and the success probability. Moreover, we have studied
how each basic operator contributes to the coherence in GSA, and
proved that the amount of operator coherence of $H_{O}$ relies on
the size of the database $N$, the success probability and the target
states. Following the idea in \cite{PMQ}, we have also discussed the
operator coherence of $|\psi_{kH_{O}}\rangle$ for different target
states. Finally, when $\alpha\in(1,2]$, we have derived the
variations of operator coherence between two consecutive iterations
of $H_{O}$, $H_{P}$ and $G$ in GSA, and the variations of
suboperator coherence of each basic operator $H_{O}$ and $H_{P}$ in
one Grover iteration. The operators $H_{O}$ and $H_{P}$ have
different effects on coherence, one produces coherence and another
depletes it. Coherence of the $H^{\otimes n}$ is not always
depleted, but depleted and produced alternatively. It oscillates
during Grover's search algorithm application.

In addition, when $\alpha\in(0,1)$ and $\alpha\in(1,2]$, the
entangled target state has different impacts on the Tsallis relative
$\alpha$ entropy of coherence. When $\alpha\in(0,1)$, the coherence
is smaller when the superposition state of targets is an entangled
one. However, when $\alpha\in(1,2]$, the coherence is larger when
the superposition state of targets is an entangled one. It is would
be interesting to study how the entanglement of the superposition
state of targets is related to the coherence quantitatively. Our
results may shed some new light on the study of the coherence
dynamics in quantum algorithms, and provide new insights into
quantum information processing tasks.

Utilizing the relative entropy of coherence and the
$l_1$ norm of coherence, it has been pointed out in \cite{LYC} that
coherence of the system states reduces to the minimum in company
with the successful implementation of Grover's algorithm. In this
paper, we can draw the same conclusion when the Tsallis relative
$\alpha$ entropy of coherence is employed. Nevertheless, similar
assertion does not hold if other quantifiers of a resource, like
quantum entanglement, are used. This peculiar character of quantum
coherence may be beneficial for designing new quantum algorithms in
the future.

\vskip0.1in

\noindent

\subsubsection*{Acknowledgements}
\small {The authors would like to express their sincere gratitude to
the anonymous referees, which greatly improved this paper. This work
was supported by National Natural Science Foundation of China (Grant
Nos. 12161056, 12075159, 12171044); Beijing Natural Science
Foundation (Grant No. Z190005); the Academician Innovation Platform
of Hainan Province.}


\subsubsection*{Competing interests}
\small {The authors declare no competing interests.}


\subsubsection*{Data availability}
\small {Data sharing not applicable to this article as no datasets
were generated or analysed during the current study.}


\end{document}